\documentclass[final,5p,times,twocolumn]{elsarticle}

\usepackage{graphicx}
\usepackage{epsfig}
\usepackage{amssymb}
\journal{Nuclear Instruments and Methods A}

\def\ff{$\phi$--factory}
\def\ifm#1{\relax\ifmmode#1\else$#1$\fi}
\def\DAF{DA\char8NE}

\def\ks{\ifm{K_S}} \def\kl{\ifm{K_L}}
  
\def\kpm{\ifm{K^\pm}}  
\def\figb#1;#2;{\parbox{#2cm}{\epsfig{file=#1.eps,width=#2cm}}}
\begin{document}
\begin{frontmatter}



\title{Status of the Cylindical-GEM project for the KLOE-2 Inner Tracker}

% if there is only one institution, use this form:
%\author{John Author, Giovanna Autore}
%\address{University of Wisdom, Physics City, Scienceland}

% else, use optional labels to link authors explicitly to addresses,
% as shown below:
\author[C]{A.~Balla}
\author[C]{G.~Bencivenni}
%\author[D]{M.~Capodiferro}
\author[C]{S.~Cerioni}
\author[C]{P.~Ciambrone} 
\author[C]{E.~De~Lucia\corref{*}\cortext[*]{Corresponding
    author}}
\ead{erika.delucia@lnf.infn.it}
\author[A]{G.~De~Robertis}
%\author[D]{A.~Di~Domenico}
\author[C]{D.~Domenici} 
%\author[C]{J.~Dong}
\author[C]{G.~Felici}  
\author[C]{M.~Gatta}  
\author[C]{M.~Jacewicz}  
\author[A]{N.~Lacalamita}
\author[C]{S.~Lauciani}
\author[A]{R.~Liuzzi}
\author[A]{F.~Loddo}
\author[A]{M.~Mongelli}
\author[B]{G.~Morello}
%\author[C]{V.~Patera}
\author[D]{A.~Pelosi}
\author[C]{M.~Pistilli}
\author[C]{L.~Quintieri}
\author[A]{A.~Ranieri}
%\author[B]{M.~Schioppa}
%\author[C]{E.~Tshadadze}  
\author[A]{V.~Valentino}
\address[A]{INFN Sezione di Bari, Bari, Italy}
\address[B]{INFN gruppo collegato di Cosenza, Cosenza, Italy}
\address[C]{Laboratori Nazionali di Frascati dell'INFN, Frascati, Italy}
\address[D]{Dipartimento di Fisica, "Sapienza" Universit\`a di Roma and
  INFN Sezione di Roma, Roma, Italy}
%%%%%%%%%%%%%%%%%%%%%%%%%%%%%%%%%%%%%%%%%%%%%%%
\begin{abstract}
The status of the R\&D on the Cylindrical-GEM (CGEM) detector foreseen as
Inner Tracker for KLOE-2, the upgrade of the KLOE experiment at the \DAF\
\ff\ , will be presented. 
The R\&D includes several activities: i) the
construction and complete characterization of the full-size CGEM prototype,
equipped with 650 $\mu$m pitch 1-D longitudinal strips; ii) the study of the
2-D readout with XV patterned strips and operation in magnetic field (up to
1.5T), performed with small planar prototypes in a dedicated test at the H4-SPS
beam facility; iii) the characterization of the single-mask GEM technology 
for the realization of large-area GEM foils. 
\end{abstract}

\end{frontmatter}
\section{Introduction}
%%%%%%%%%%%%%%%%%%%%%%%%%%%%%%%%%%%%%%%%%%%%%%%%%%%%%%%%%%%
After the completion of the KLOE data taking~\cite{bib:rivista},
the proposal of a new run  with an
upgraded KLOE detector, KLOE-2~\cite{bib:kloe2}, at an upgraded \DAF\ 
 machine has been accepted and will start in spring 2010~\cite{bib:rollin}.
The KLOE-2 physics program will be focused on physics coming from the
interaction point (IP), where the $\phi$-meson is produced: 
 \ks\ , \kpm\ , $\eta$ and $\eta^{'}$ decays as well as \ks\-\kl\
interference  and search for physics beyond the Standard Model.
The improvement in the reconstruction performance for tracks near the
interaction region is then of fundamental importance for the accomplishment of this physics program.
\par
\begin{figure}[!h]
\centering
\includegraphics[width=0.4\textwidth]{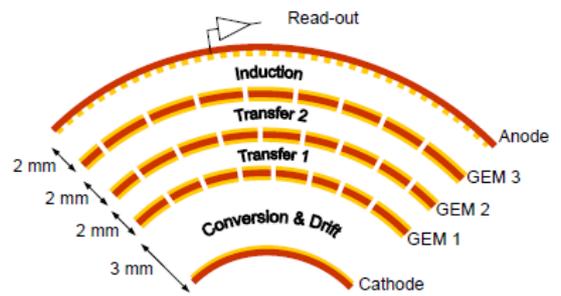}
\caption{Cross-section of the triple-GEM detector.}
\label{fig:gemscheme}
\end{figure}

 After a first phase with the installation of the low-energy $e^+e^-$
(LET)~\cite{bib:let} and  high-energy $e^+e^-$ (HET)~\cite{bib:let_het} tagging
systems for the identification and study of $\gamma-\gamma$ events,
the detector will be upgraded with the insertion of an
Inner Tracker (IT) between the beam pipe and the Drift 
Chamber (DC) inner wall. A crucial design parameter is the resolution on 
the \ks\ decay point, occurring within few cm from the IP. 
An accurate study on quantum interferometry measurement showed that an improvement on 
this resolution of about a factor of 3 with respect to the present 
value ($\simeq$~6~mm) is required~\cite{bib:it_tdr}. 
The IT contribution  to the overall material budget
has to be carefully taken into account in order to 
minimize multiple scattering contribution to the track momentum resolution and photon conversions
before the DC volume. 
The main requirements for the IT can be summarized as: 
\begin{enumerate}[i.]
\item $\sigma_{\rm r\phi}\sim 200~\mu$m and $\sigma_{\rm z} \sim 500~\mu$m spatial
  resolutions;
\item total material budget below 2$\%$ of the radiation length X$_0$;
%  to minimize the multiple scattering effects and photon conversion before the DC volume;
\item 5 kHz/cm$^2$ rate capability.
\end{enumerate}
The adopted solution is the Cylindrical-GEM (CGEM)~\cite{bib:it_tdr}, a triple-GEM detector composed by concentric cylindrical electrodes (fig.~\ref{fig:gemscheme}): Cathode, 3 GEM foils for the multiplication stage (gain $\sim$10$^4$) and Anode, acting also as the readout circuit. 
The high rate capability of the GEM (up to 1~MHz/mm$^2$~\cite{marcopoli}) makes this detectors suitable to be placed near the interaction point of a high-luminosity collider machine.
The IT will be then composed by 5 concentric CGEM detection layers at radii
from 13 cm, to preserve the \ks-\kl\ quantum 
interference region, to 23 cm due to the constrained from DC inner wall at 25 cm.
The total active length for all layers will be 70 cm. The anode readout of
each CGEM will be segmented  with 650 $\mu$m pitch XV patterned strips with a stereo
angle of $\sim 40^\circ$, for a total of about 30,000 FEE
channels. The Front-End Electronics for the IT is based on the
new GASTONE ASIC~\cite{bib:gastone}, a 64 channels chip composed by four different stages: a charge preamplifier with 20mV/fC sensitivity, a shaper, a leading-edge
discriminator with a programmable threshold and a monostable stretcher of
the digital signal, to synchronize with the KLOE Level1 trigger.
\par In the following the main stages of the R\&D project~\cite{bib:it_tdr} will be discussed: i) the construction and complete characterization of a full-scale CGEM prototype, ii) the study of the XV strip readout configuration and its operation in magnetic field and iii) the construction and characterization of a large area GEM realized with the new single-mask photolitografic technique.
%%%%%%%%%%%%%%%%%%%%%%%%%%%%%%%%%%%%%%%%%%%%%%%%%%%%%%%%%%%
\section{Cylindrical-GEM prototype}
%%%%%%%%%%%%%%%%%%%%%%%%%%%%%%%%%%%%%%%%%%%%%%%%%%%%%%%%%%%
 A full-scale CGEM prototype has been built with 15 cm radius and
35 cm active length~\cite{bib:cgem}. 
The cylindrical electrodes have been obtained from very light polyimide foils (50 $\mu$m thick 
for GEMs, 100 $\mu$m for Cathode, 100 $\mu$m for Anode) rolled onto machined PTFE cylinders, acting as molds, and then glued exploiting the vacuum bag technique.
The cylinders are then inserted one into the other and fiberglass annular flanges
are glued at the ends, acting as spacers for the gaps and as supporting
mechanics~\cite{bib:cgem_mech}. 
The CGEM is therefore a low-mass, fully cylindrical and dead-zone free GEM based detector with no support frames required inside its active area. 
The anode was patterned with 650 $\mu$m pitch
longitudinal strips, reconstructing the r-$\phi$ coordinate, for a total of
1538 axial strips. The chamber, tested with X-rays to check the uniformity over the large
surface, has been then equipped with the 16-channels prototype of GASTONE~\cite{bib:gastone} and
tested at the CERN PS with a 10 GeV pion beam. The results show the
expected spatial resolution from a digital readout of 650 $\mu$m pitch
strips~\cite{bib:cgem_res} and a 99.6$\%$ efficiency. 
%\par The construction of the CGEM prototype, its safe operation and
%extensive test validated this innovative detector idea.
%%%%%%%%%%%%%%%%%%%%%%%%%%%%%%%%%%%%%%%%%%%%%%%%%%%%%%%%%%%
\section{XV readout and operation in magnetic field}
\begin{figure}[!htb]
\centering
\includegraphics[width=0.3\textwidth]{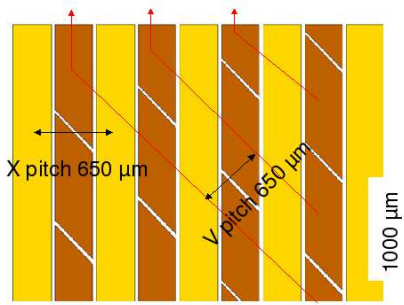}
\caption{Scheme of the XV readout configuration.}
\label{2D_readout1}
\end{figure}

%%%%%%%%%%%%%%%%%%%%%%%%%%%%%%%%%%%%%%%%%%%%%%%%%%%%%%%%%%%
 A typical orthogonal XY readout can not be used for the inner tracker,
due to its cylindrical geometry. The final IT readout will be then
performed with an XV pattern of strips and pads engraved on a polyimide
foil substrate, 100 $\mu$m thick (fig.~\ref{2D_readout1}). 
 The X strips with 650 $\mu$m pitch will provide the r-$\phi$
coordinate while the pads, connected through internal vias to form V strips
with 650$\mu$m pitch, will provide the z coordinate. This quite innovative
readout solution was not implemented on the CGEM prototype, therefore its
characteristics have been extensively studied with dedicated planar
chambers. In addition, since the IT will operate inside KLOE's
 magnetic field, the effects on the cluster  
formation and electronics readout had to be studied. 
\par
\begin{figure}[htb]
\centering
\includegraphics[width=3.0in]{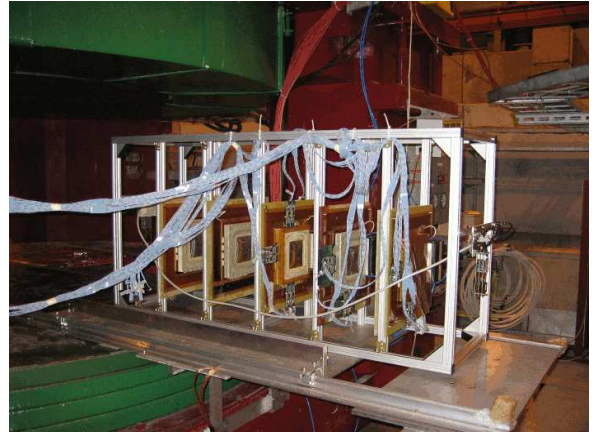}
\caption{Setup of the test beam at CERN with planar chambers with
       2D-readout.}KLOE-2 inner tracker technical design report
\label{setup_xv_testbeam}
\end{figure}

 To address these issues a dedicated test has been done 
at the H4 permanent facility, setup at the CERN-SPS 150GeV 
pion beam line within the RD51 Collaboration~\cite{bib:rd51}.
Five 10x10 cm$^2$ planar triple-GEM (PGEM) detectors with 650 $\mu$m pitch readout
have been assembled and succesfully tested: four chambers with standard XY
readout and the fifth with the XV readout under investigation.  
The setup was 1 meter long with detectors placed equidistantly with the XV
chamber placed in the center (fig.~\ref{setup_xv_testbeam}).
For the operation in magnetic field, the GOLIATH magnet was used, providing a field
adjustable up to 1.5 T perpendicular to the horizontal beam-plane.
To fully cover the area illuminated by the SPS beam,
the planar chambers were partially equipped with 22 digital readout GASTONE boards,
32 channels each, four on each XY chamber and six on the XV chamber.
The coincidence of 6 scintillators (3x3  cm$^2$) readout by silicon 
photomultipliers, three upstream and three downstream, provided the trigger signal for the acquisition.
The same working point of the CGEM prototype has been used:
Ar/CO2 (70/30) gas mixture and operating voltages V$_{\rm fields}$ = 1.5/3/3/5 kV/cm and
 V$_{\rm GEM}$ = 390/380/370 V ($\sum$V$_G$ = 1140V, Gain=2$\times$10$^4$). The GASTONE threshold was
 set at 3.5 fC. 
\par
\begin{figure}[!h]
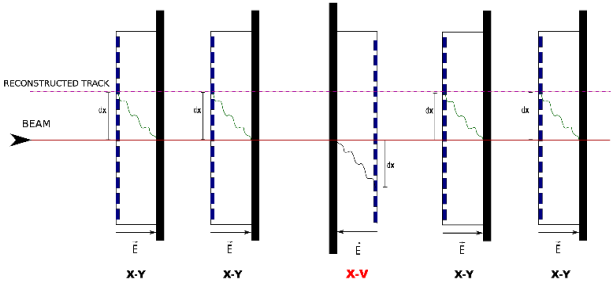

\centering
\figb setup_lorentz_angle;8.;
\caption{Test beam setup and definition of the measured quantity.}
\label{fig:lorentz2}
\end{figure}

\begin{figure}[!h]
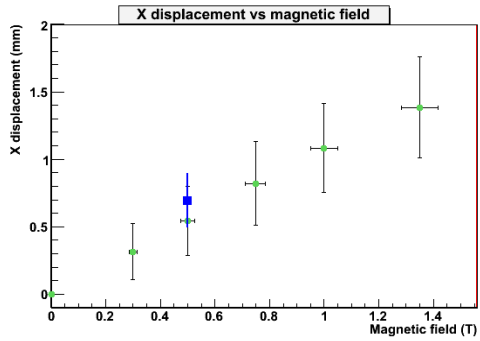

\centering
\figb lorentz_angle_bfield;7.;\\
%\figb PGEMpics/ ;7.;\kern0.2cm \figb PGEMpics/ ;5.;
\caption{Displacement dx as a function of the magnetic field (green points) with the
  result from GARFIELD simulation at B=0.5 T (blue
  square).}
\label{fig:lorentz_bfield}
\end{figure}

\begin{figure}[!h]
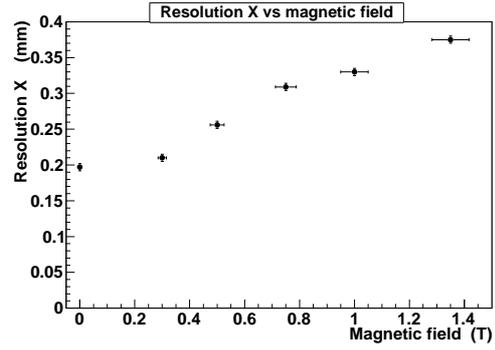

\centering
\figb xres2;7;
\caption{Resolution on the X coordinate as a function of the magnetic field.}
\label{fig:resolux}
\end{figure}

\begin{figure}[!h]
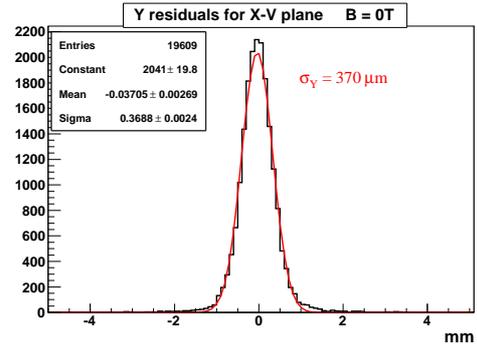

\centering
\figb vres2;7.;
\caption{ 
  Resolution on the Y coordinate at B=0 T.}
\label{fig:resoluy}
\end{figure}

\begin{figure}[!h]
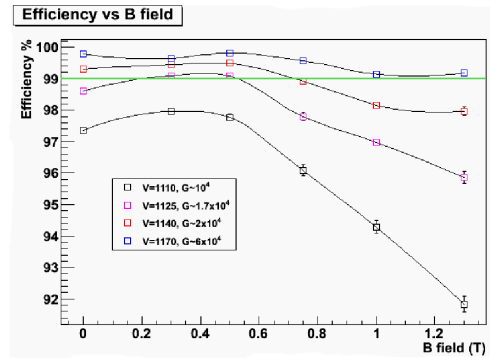

\centering
\figb efficiency_vs_Bfield_Gains;7.;
\caption{ Efficiency as function of the magnetic field obtained operating
  the XV chamber with four different gain values. }
\label{fig:pgemeff}
\end{figure}

 The effect of the magnetic field (B) is twofold: a displacement dx and a
spread $\sigma_{\rm dx}$ of the charge over the readout plane. 
The expected values obtained from simulation studies of our chambers done with GARFIELD
are dx=600~$\mu$m and $\sigma_{\rm dx}$=200~$\mu$m at B=0.5 T~\cite{bib:it_tdr}.   
In the test beam configuration the magnetic field effect was mainly present on the X-view. 
The setup used to measure the displacement on the XV chamber due to the
magnetic field is shown in fig.~\ref{fig:lorentz2}.
All four XY chambers are likewise oriented, with the
same anode-cathode configuration, and provide the external tracking system for the XV
chamber which is instead in a cathode-anode arrangement, reversed with respect to
the other chambers.  
Since the XY chambers are subject to the same Lorentz force, the reconstructed
track will be shifted by the same offset dx with respect to the true
track trajectory. The displacement in the XV chamber instead will be of the
same magnitude dx but with opposite direction, due to the reversed cathode-anode arrangement. 
First, with zero magnetic field (B=0 T), the setup was aligned  
to a few micrometer precision and then, with the magnetic field turned on, 
the total displacement (D) between the track reconstructed by the XY telescope
and the point in the XV chamber was measured: D = 2$\times$dx (fig.~\ref{fig:lorentz2}).
%During KLOE data taking the magnetic field was set at 0.52 T.  
%To improve the acceptance for low momentum tracks, the option with a lower
%magnetic field value, e.g. 0.3 T, is under study for the KLOE2 project.
The displacement dx was measured for 5 values of the magnetic field
and found in good agreement with the value obtained from the GARFIELD simulation at B=0.5 T (fig.~\ref{fig:lorentz_bfield}). 
%Such effect will be properly taken into account in the reconstruction procedure. 
To study the XV chamber performance, we measured the resolution in
both X and Y coordinates: the X coordinate is measured
directly from the X strips while the Y coordinate is obtained from the
crossing of both X and V strip readout.
Fig.~\ref{fig:resolux} shows the resolution on the X coordinate as a
function of the magnetic field, the values ranging from 200 $\mu$m at B=0
T up to 380 $\mu$m at B=1.35 T. 
The resolution on the Y coordinate measurement is $\sim$370 $\mu$m at B=0 T,
in agreement with what expected from the digital readout of the two X and V
views (fig.~\ref{fig:resoluy}).
The performance of the front-end chip GASTONE have been studied measuring the cluster size and 
reconstruction efficiency, defined as the presence
of a cluster in the XV chamber when a candidate track was reconstructed using four XY
chambers. Fig.~\ref{fig:pgemeff} shows the reconstruction efficiency measured as a function of the magnetic
field and  operating the XV chamber with four different gain values.
The efficiency for the nominal KLOE magnetic field B=0.52 T and voltage settings was measured to
exceed 99$\%$, slightly decreasing at higher B fields. In fact the increase of the magnetic field
leads to a larger spread of the charge over the readout strips and causes a
reduction of the charge seen by each single pre-amplifier channel, with a
consequent efficiency drop. Thus the increase of the magnetic field
requires for higher gains to efficiently operate the chamber. The variation
of the magnetic fields within the KLOE-2 planned values (0.3-0.5 T), to improve the acceptance for low momentum tracks, has a negligible effect on the reconstruction efficiency in the voltage range around our working point.
The charge sharing, grounding and cross-talk between strips could have been in
principle different for X and V views, due to the different readout geometry. 
Our measurements demonstrated good behavior of both X and V readout views, 
which appear to have equalized response. 
%($V_{ref}$= 1140 V).
%%%%%%%%%%%%%%%%%%%%%%%%%%%%%%%%%%%%%%%%%%%%%%%%%%%%%%%%%%%
\section{Large Area GEM}
%%%%%%%%%%%%%%%%%%%%%%%%%%%%%%%%%%%%%%%%%%%%%%%%%%%%%%%%%%%
To build the IT outermost layer, a GEM foil as large as 1440x700
mm$^2$ is needed. This foil can be obtained splicing 3 separate 480x700
mm$^2$ foils with a technique that we have developed, using an epoxy
adhesive and a vacuum bag. The urge for larger GEM foils has driven a
change of the production procedure by CERN TS-DEM-PMT laboratory, switching to a
single-mask etching, more suitable for a large surface~\cite{bib:mvilla}.
The new GEMs have quasi-cylindrical holes and a new characterization is
necessary. In order to check the uniformity of
 the new single-mask GEMs over a large area, we will build a 700x300 mm$^2$
 planar triple-GEM. Dedicated tools for the stretching, handling and
 assembling of such large foils have been designed and realized.
% ANSYS simulations indicate that even on such a large area, with a tension
% of 1~kg/cm the maximum sag due to combined gravitational and electrostatic
% effects is only 20~$\mu$m. 
 The chamber, that will be the largest GEM detector ever operated, will be
 equipped with the GASTONE 64-channels final release and readout with
 the Off Gastone Electronic (OGE) Board~\cite{bib:it_tdr}. 
A dedicated test beam of the Large Area GEM is foreseen this year at the
CERN. The external tracking system will be provided by the four XY
 chambers used for the XV readout studies, replacing the XV chamber with
 the Large Area GEM.
%%%%%%%%%%%%%%%%%%%%%%%%%%%%%%%%%%%%%%%%%%%%%%%%%%%%%%%%%%%
\section{Finalizing the project: 3-D Finite Element simulation}
%%%%%%%%%%%%%%%%%%%%%%%%%%%%%%%%%%%%%%%%%%%%%%%%%%%%%%%%%%%
ANSYS~\cite{bib:ansys} finite element 3-D simulation of the CGEM has been developed
to estimate the structural response under tensile loads: induced strain,
stress and displacements~\cite{bib:mech_simul}. A detailed characterization of the materials has
been done to implement an accurate description of the mechanical
behavior. The model has been then validated by comparison with measurements
done with CGEM and PGEM prototypes.
\begin{figure}[hbt] 
\centering 
\includegraphics[width=0.4\textwidth]{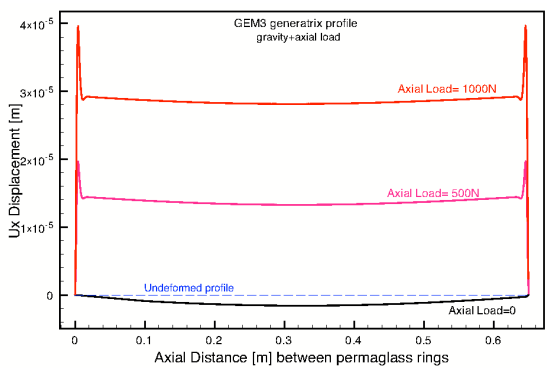}
\caption{ANSYS simulation of the 700 mm length CGEM: radial displacement of
  the generatrix profile of a GEM foil obtained applying several tensile loads.}
\label{fig:ansys}
\end{figure}
Fig.~\ref{fig:ansys} shows the results
obtained with the simulation of the 700 mm length CGEM for the generatrix
profile of a GEM foil, with gravitational effect only and once an axial load is applied to improve the stability of the foils.
The gravitational sag is $O(\mu m)$ therefore negligible and the maximum value of the expected 
radial displacement is 20 $\mu$m for 500 N axial load and becomes 40 $\mu$m
with 1000 N. The peculiar effect observed on the edges of the distributions
is due to the presence on the GEM foil edges of 5 mm of kapton connecting
two more rigid structures radially displaced: the GEM active zone (two-side
copper-clad kapton foil) and the Permaglass ring. This originates an S-shaped deformation which translates into the observed egde effect. Besides this, the radial displacement is constant in the GEM active zone and similar distributions have been obtained for Cathode and Anode showing that the structure of the CGEM gaps is preserved.
%%%%%%%%%%%%%%%%%%%%%%%%%%%%%%%%%%%%%%%%%%%%%%%%%%%%%%%%%%%
\section{Conclusions}
%%%%%%%%%%%%%%%%%%%%%%%%%%%%%%%%%%%%%%%%%%%%%%%%%%%%%%%%%%%
The construction, safe operation and extensive test of an almost full-size Cylindrical-GEM prototype has
demonstrated the feasibility of such a novel low-mass and dead-zone-free
vertex detector. The final readout configuration has been validated with
the successful test of the small planar prototypes operated in magnetic
field: very good results in terms of spatial resolution, efficiency and
cluster size have been obtained. A large planar GEM prototype will be built using foils
realized with the new single-mask technique to test their quality and homogeneity together with the GASTONE 64-channels readout final release.
The R\&D phase on Cylindrical-GEM is practically concluded. The
project of the KLOE-2 Inner Tracker has been recently approved and its construction will start this year in order to be ready for insertion in the KLOE
detector by summer 2011.

%\section{}
%\label{}



\end{document}
\endinput